# PECULIARITIES OF THE SERS SPECTRA OF THE HYDROQUINONE MOLECULE ADSORBED ON TITANIUM DIOXID

A.M. Polubotko[*], V.P. Chelibanov[**]

*A.F. Ioffe Physico-Technical Institute Russian Academy of Sciences,

Politechnicheskaya 26, 194021 Saint Petersburg Russia, Tel: (812) 274-77-29,

Fax: (812) 297-10-17  E-mail: alex.marina@mail.ioffe.ru

[1]State University of Information Technologies, Mechanics and Optics,

Kronverkskii 49, 197101 Saint Petersburg, RUSSIA  E-mail:

Chelibanov@gmail.com

## Abstract

The SERS spectrum of hydroquinone, adsorbed on nanoparticles of titanium dioxide ($TiO_2$) is analyzed. It is pointed out that the enhancement is stronger for larger mean size of nanoparticles that is in an agreement with the electrostatic approximation. In addition it is found that there are the lines, which are forbidden in usual Raman spectra. Along with this there is the enhancement, caused both by the normal and tangential components of the electric field. This result is in agreement with the theory of SERS on semiconductor and dielectric substrates. Discovery of the forbidden lines indicates sufficiently large role of the strong quadrupole light-molecule interaction in such a system.

The study of the SERS phenomenon on semiconductor and dielectric substrates is of a significant interest both from experimental and theoretical points of view. In [1] it was shown that the reason of SERS in this case is the surface roughness, such as for metal. The enhancement arises in small regions of the surface with very large positive curvature. In [1] it was demonstrated that the enhancement on dielectric and semiconductor substrates is less than on metals with the same value of the modulus of the dielectric constant. This result is associated with the fact that semiconductors and dielectrics are transparent for the electromagnetic field in a wide range of frequencies, while the metal is not transparent and tends push out the field. Therefore the systems with the semiconductors and dielectrics have "less inhomogeneity" of the medium, compared with the metal that results in a weaker enhancement of the field and its derivatives. However, in accordance with experimental and theoretical results of [1], there must the enhancement both of the normal and tangential components of the electric field compared with metal that results in some features of the SERS spectra. We investigated the spectra of hydroquinone, adsorbed on titanium dioxide ($TiO_2$). As it is well known from literature, hydroquinone is a symmetrical molecule with the $C_{2h}$ symmetry group (Figure 1). In accordance with the ideas expounded in [2], it can form a molecular crystal of triclinic and monoclinic syngony. The hydroquinone molecules form chains and connect to each other via the hydrogen atoms. Investigation of the Raman and infrared spectra demonstrated [2] that the vibrational frequencies of the hydroquinone molecules are very close for both cases and are close to the frequencies in a vapor or in a solution of $CH_3CN$. This result allows us to use the assignment to the irreducible representations of the symmetry group of hydroquinone, obtained in [3-5] (Table 1).

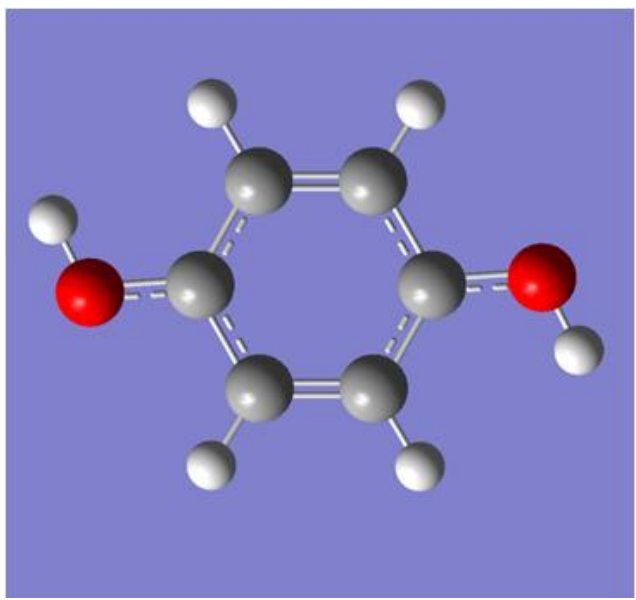

Figure 1. The hydroquinone molecule.

Table 1. Assignment of the hydroquinone lines in the SERS spectra for nanoparticles of $TiO_2$ with mean sizes 10 and 80 nm. (vw-very weak, w-weak, m-middle, s-strong, sh-shoulder)

| Hydroquinone on $TiO_2$. SERS. The mean size of the particles 10 nm. | Hydroquinone on $TiO_2$. SERS. The mean size of the particles 80 nm. | The lines of pure $TiO_2$ | Irreducible representations. ($C_{2h}$ symmetry group). |
|---|---|---|---|
| 196 vw. | | | $A_u$ |
| 376 vw. | | | $B_g$ |
| | | 397-400 | |
| | | 476 | |
| | | 515-517 | |

|        |              | 638-641 |         |
|--------|--------------|---------|---------|
| 704 vw. | 704 vw.     |         | $B_g$   |
| 808 w.  | .           |         | absent  |
| 811 w.  | 811 m.      |         | absent  |
| 843     | 843 s       |         | $A_g$   |
| 853 vw. | 853 s       |         | $A_g$   |
|         | 894 vw..    |         | absent  |
|         | 915 vw.     |         | absent  |
|         | 1001 w.     |         | $B_u$   |
| 1153 sh.| 1153 vw.    |         | $A_g$   |
| 1159 w. | 1159        |         | $A_g$   |
|         | 1220 vw.    |         | $B_u$   |
|         | 1241 vw.    |         | $B_u$   |
|         | 1263 s.     |         | $A_g$   |
| 1269    | 1267 s. sh. |         | $A_g$   |
| 1274    | 1274 sh.    |         | $A_g$   |
| 1280    |             |         | absent  |
| 1185    |             |         | absent  |
| 1500 w. | 1500 w.     |         | $B_u$   |
|         | 1590 vw. sh.|         | $A_g$   |
| 1607 w..| 1607 m.     |         | $A_g$   |

The spectra of hydroquinone, adsorbed on colloidal particles of $TiO_2$ with a mean sizes 10 and 80 nm within the wavenumbers interval of 600-1700 $cm^{-1}$ are shown on Figure 2. One should note that the spectra were taken at a wavelength of the incident light 785 nm. The enhancement coefficient was $\sim 10^3 - 10^4$. One can find the used spectrometer description in [6,7].

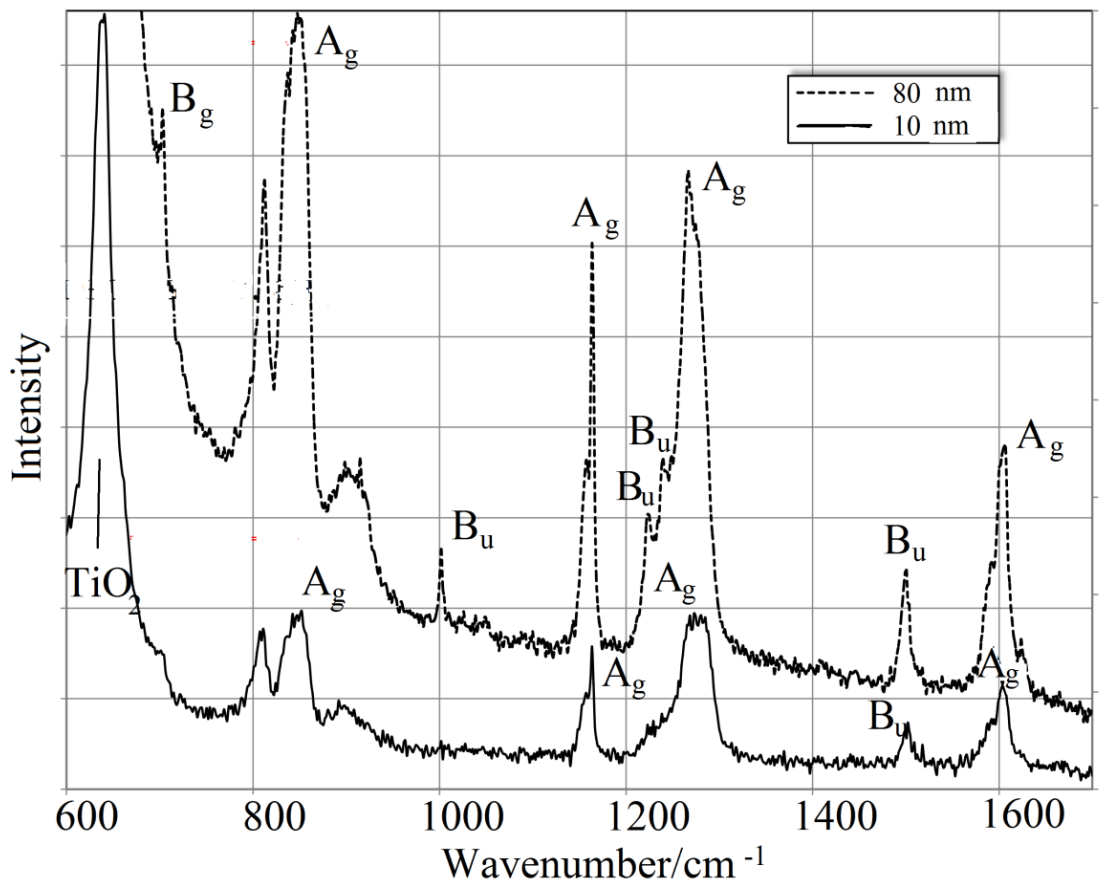

Figure 2. The SERS spectrum of hydroquinone, adsorbed on nanoparticles of $TiO_2$ with the mean sizes 10 and 80 nm in the range of wavenumbers 600-1700 $cm^{-1}$.

In accordance with our ideas, the hydroquinone molecule adsorbs parallel to the surface of nanoparticles. The peculiarity of the spectrum taken off from the particles with a mean size 10 nm is appearance of doublets in the lines with the wavenumbers (808,811) and (1153,1159) $cm^{-1}$, and also sufficiently broad bands and their fine structure in the region of 849 and 1270 $cm^{-1}$. Sufficiently broad width and a fine structure of the bands apparently indicates sufficiently strong interaction of molecules with the substrate and existence of non equivalent positions of the molecules. For the spectrum taken off for the nanoparticles with the size 10 nm, practically all lines refer to the vibrations with the unit irreducible representation $A_g$, such as in usual Raman scattering. However, one line with the wavenumber ~1500-1512 $cm^{-1}$ refers to the vibration with the irreducible representation $B_u$, which describes transformational properties of the dipole moment components $d_x$ and $d_y$, which are parallel to the surface. Appearance of this line, which is forbidden in usual Raman scattering is associated with existence of sufficiently strong quadrupole light-molecule interaction in this system. Its weak intensity indicates that the quadrupole interaction is sufficiently weak in this case, compared with the case of a metal. In addition, appearance of the line, which refer to the irreducible representation $B_u$, points out validity of our theoretical result [1], which indicates possibility of enhancement not only of the normal component of the electric field, but on the enhancement of the tangential components as opposite to the case of a metal, where such enhancement is very weak, or absent at all, because of its very large conductance. In case of large particles, with a mean size ~80 nm the SERS spectrum is enhanced significantly stronger. This result one can explain by the fact that in the area, where the diffraction on nanoparticles can be described within the framework of a quasi-static approximation for the electric field, the scattering intensity must be stronger for the particles of larger size. In this case there are several additional weak forbidden lines, which are assigned to the vibrations with the irreducible representation $B_u$ with the wavenumbers 1001, 1220 and 1241 $cm^{-1}$. In addition there is an additional line at 704 $cm^{-1}$, associated with the vibration, which refers to the irreducible representation $B_g$.

Within the range of the wavenumbers 100-700 $cm^{-1}$ (Figure 3), the spectrum of adsorbed hydroquinone is masked by the overlapping by the strong lines, which refer

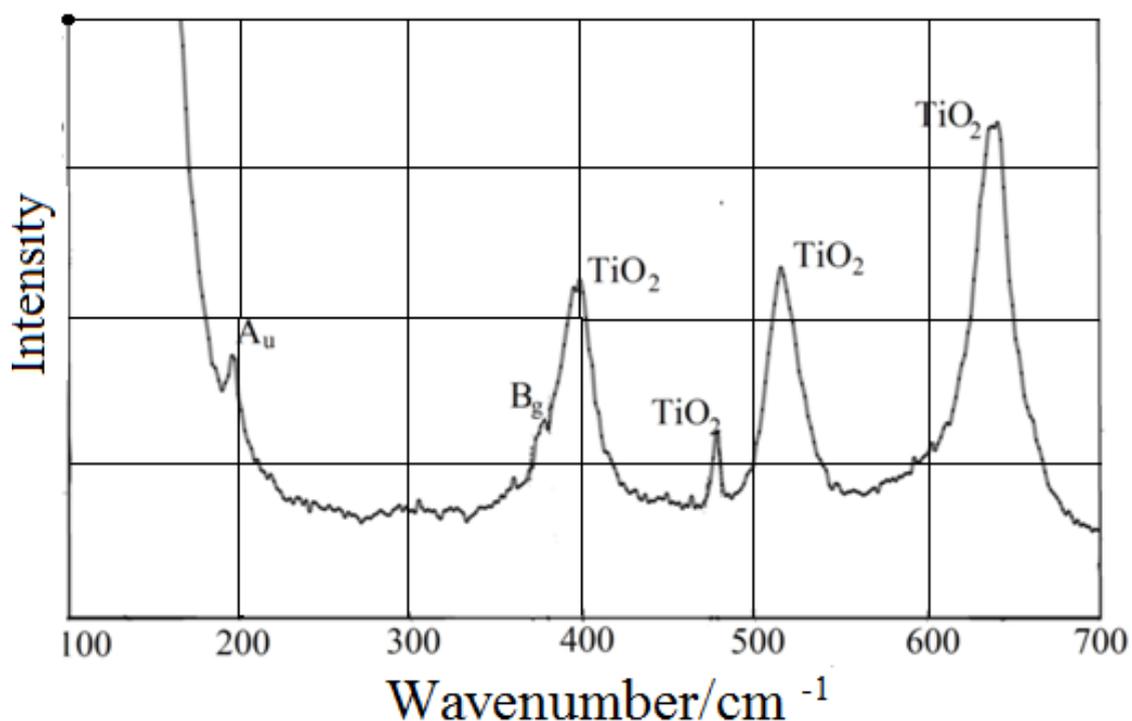

Figure 3. The SERS spectrum of hydroquinone, adsorbed on nanoparticles with a mean size 10 nm within the range of the wavenumbers 100-700 $cm^{-1}$.

to the vibrations of the lattice of $TiO_2$ with the wavenumbers 397-400, 476, 512-517 and 638 $cm^{-1}$. However, one can see in this region a weak forbidden line with the wavenumber 196 $cm^{-1}$, which refers to the vibration with the irreducible representation $A_u$ and the line with the irreducible representation $B_g$ at 378 $cm^{-1}$

In general, investigation of the hydroquinone spectrum points out the correctness of our SERS theory on semiconductor and dielectric substrates [1], and on the necessity to take into account the quadrupole interaction, which is sufficiently strong in this system and results in appearance of weak forbidden lines. One should note that in literature on SERS, SEHRS, SEIRA and other surface enhanced processes the strong quadrupole light-molecule interaction usually is named as a gradient field mechanism, [8, 9] for example. It is necessary to note that the use of such terminology is a crude mistake. The concept of gradient is well defined in mathematics and is

$$\text{grad}\,\Psi = i\frac{\partial \Psi}{\partial x} + j\frac{\partial \Psi}{\partial y} + k\frac{\partial \Psi}{\partial z} \quad ,$$

where $\Psi$ is a scalar field, and cannot be used for the vector field that is made by the above authors. Here $i, j, k$ are the unit vectors along $x, y, z$ axes. One can find this definition in any textbook on higher mathematics, [10] for example.